# Highly inhomogeneous interactions between background climate and urban warming across typical local climate zones in heatwave and non-heatwave days

Jing Kong[1], Yongling Zhao[2], Kai Gao[3], Dominik Strebel[2], Jan Carmeliet[2], Chengwang Lei[1]

1. Centre of Wind, Waves and Water, The University of Sydney, Australia

2. Department of Mechanical and Process Engineering, ETH Zürich, Zürich, Switzerland

3. Institute of Future Cities, Chinese University of Hong Kong, Hong Kong, China

## Abstract

Urban heat island (UHI) in conjunction with heatwave (HW) leads to exacerbation of thermal stress in urban areas. Prior research on UHI and HW has predominantly concentrated on examining the thermal conditions at the surface and near-surface, with few investigations extending to the radiative and dynamical interactions of UHI and HW, particularly with a focus on the inhomogeneities across local climate zones (LCZs). Here, we analyse the temperature disparity between HW and non-HW conditions across LCZs in the Sydney area by quantifying the contributions of individual radiative and dynamical processes using the coupled surface-atmosphere climate feedback-response analysis method (CFRAM). Three HW events in 2017, 2019, and 2020 are simulated using the Weather Research and Forecasting (WRF) model coupled with the Single-Layer Urban Canopy Model (SLUCM). The maximum temperature difference between HW and non-HW days may reach up to 10 K, with the increased net solar radiation during HWs being comparable to the typical level of anthropogenic heat flux in urban areas. It is also found that the reduction of clouds, the presence of vapor, and the increase of sensible heat contribute to the warming effect at different levels, with the contribution of clouds being the most dominant. Conversely, the generation of dry convection and the increase of latent heat flux lead to mitigating effects, with the latter being more dominant and capable of causing up to 10 K surface temperature difference between LCZ1 (compact high-rise) and LCZ9 (sparsely built). The differences in the contributions of climate feedback processes across different LCZs become more evident during more severe and humid HWs. These findings underscore the necessity of implementing local climate zone-tailored heat mitigation strategies.

1. Introduction

In the context of climate change and global warming, heatwaves (HWs) have emerged as a significant meteorological danger due to their adverse impact on public health [1, 2], socioeconomic development [3], agriculture [4], wildfire frequency and intensity [5], and ecosystems [6]. Extreme heat-related risks have been reported in several regions, including Canada and the United States [7], the United Kingdom [8], China [9], Australia [10], and Europe [11]. There exists a prevailing consensus that the frequency, intensity, and duration of HWs have increased and will continue to increase due to increasing greenhouse gas concentrations at both regional and global scales [12-14].

A HW is commonly characterised as an extended duration of exceptionally high temperature and is usually linked to a high-pressure synoptic system caused by blocking Rossby waves [15]. Two distinct categories of HWs are identified by the World Meteorological Organisation: dry HWs and moist HWs [16]. Dry HWs are characterized by stable atmospheric conditions, clear skies, high temperature, and low humidity. In contrast, moist HWs coincide with high temperature, high humidity, and are often accompanied by nocturnal cloud cover [17]. Moist HWs typically exhibit a greater release of latent heat and enhanced downward solar radiation due to anomalous subsidence [18]. The additional moisture is mostly due to abundant soil moisture and/or intensified moisture advection, such as that coming from an adjacent ocean [19, 20]. Ha and Yun [21] highlighted that increased surface water vapour plays a crucial role in trapping more heat from the surface during the night. The combined high temperature and high humidity are more hazardous to human health than dry and hot conditions due to reduced efficiency of human perspiration [22]. It is anticipated that the atmosphere will experience increased moisture as a result of global warming [23], and the intensity, frequency, and duration of moist HW events will increase [16].

Most prior research has examined the mechanisms of HW by focusing on dynamic processes linked to large-scale movements of the atmosphere, which cannot fully account for the persistence of HWs caused by high-pressure systems occurring over much shorter periods [24]. A comprehensive understanding of quantitative attributes of each atmospheric process involved in different HWs are still lacking. For studies of urban HWs, most research has examined the surface and near-surface thermal conditions within the urban canopy layer (UCL) [25-28], which refers to the layer beneath the average height of buildings and trees. Few studies [29-32] have investigated the response of the planetary boundary layer (PBL) to HWs or studied the impact of LCZs during HWs. Some studies considered only a single HW event. A quantitative examination of underlying physical processes inside the PBL is still missing. As reported by Davy and Esau [33], the depth of PBL determines the effective heat capacity of the atmosphere. How the PBL responds to different HWs is still a significant and unresolved research question.

To address these gaps, we employ the coupled atmosphere-surface climate feedback-response analysis method (CFRAM) [34], which is based on energy balance within the atmosphere-surface column, to quantitatively assess the attribution of climate feedbacks to temperature anomalies in the PBL during three moist HW events in Sydney. CFRAM uses the correlation between infrared radiation and temperature over the entire atmosphere-surface column, thus the temperature variations resulting from perturbations in external energy flux terms or feedbacks can be determined by obtaining the temperature-induced alteration of infrared radiation [34]. The effectiveness of CFRAM for decomposing radiative and non-radiative processes and assessing key processes that regulate surface and atmospheric temperatures has been demonstrated in many studies [16, 24, 35-43].

This study examines three moist HW events in the Sydney area. We aim to: (1) quantify the modulation of heat flux between the surface and atmosphere in moist HWs; and (2) comprehend primary factors that contribute to the alterations in temperature anomalies caused by HWs in different local climate zones (LCZs). The rest of this paper is organised as follows: section 2 describes the data and method, section 3 presents the main results in different LCZs and discusses the dominant feedback and limitations, and section 4 summarises the present study and discusses future work.

## 2. Data and Methods

### 2.1 Study Area

The Sydney metropolitan area is located on the east coast of Australia at around 33.9 °S latitude with elevation ranging from 0 to 500 m [44]. It is the largest city in Australia and has a population density of 8176 persons per square kilometre in the City of Sydney area and 429 persons per square kilometre in the Greater Sydney Area [45]. The climate of Sydney is categorised as a humid subtropical climate (Cfa) with warm summers and cool winters according to the Köppen-Geiger climatic classification [46]. During the summer season, the monthly averaged daily maximum temperatures observed at Observatory Hill, located in the central business district (CBD), exhibit a range between 25.2 °C and 26.0 °C [47].

### 2.2 WRF Simulation Design

The Weather Research and Forecasting (WRF) model V4.3 [48], which is a fully compressible and non-hydrostatic regional climate model, is adopted to study the impacts of HW on UHI in Sydney. The configuration of the WRF input domains is shown in Figure 1. The computational model consists of three one-way nested domains at 4.5 km resolution (D01: 181 × 133 grid cells), 1.5 km resolution (D02: 181 × 133 grid cells), and 0.5 km resolution (D03: 181 × 133 grid cells), respectively. There are 65 vertical levels from the surface up to 50hPa, with denser levels near the surface to better resolve near-surface processes. The model simulation periods are from January 25, 00:00 to February 25, 21:00, 2017 (UTC), January 20, 00:00 to February 20, 21:00, 2019 (UTC), and January 20, 00:00 to February 20, 21:00, 2020 (UTC) with the first day as the spin-up period to mitigate any inconsistency that may arise between internal dynamics and external forces. The simulated results are exported on an hourly basis.

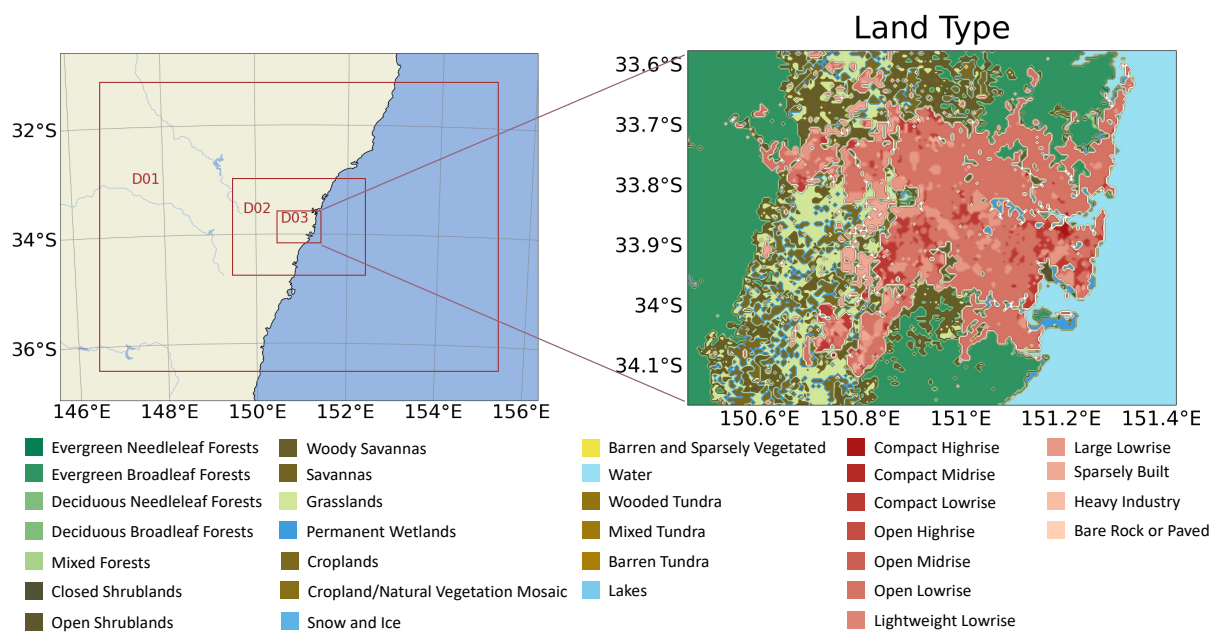

Figure 1. Configuration of the WRF input domains and land classifications in D03.

The numerical schemes adopted in this study include: the Rapid Radiative Transfer Model for GCMs (Global Circulation Models) (RRTMG) scheme [49] for longwave and shortwave radiation, the Mellor-Yamada-Janjić (MYJ) scheme [50] for Planetary Boundary Layer (PBL) modelling, which utilises the 1.5-order closure turbulence model of Mellor and Yamada [51] to represent the turbulence above the surface, the Monin-Obukhov (Janjic Eta) similarity scheme [52] for surface layer parameterization, the new Tiedke scheme [53, 54] for cumulus, the Thompson et al. scheme [55] for microphysics, and the Noah land surface model [56] coupled with the Single-Layer Urban Canopy Model (SLUCM) [57] for land surface parameterization and urban processes, respectively. The effectiveness of these schemes has been demonstrated previously for the Sydney region [25, 58].

The initial and boundary meteorological conditions are extracted from the 3-hourly ECMWF Reanalysis v5 (ERA5) model level data on a 0.25° × 0.25° grid. The time-varying sea surface temperature (SST) is applied to enhance the modelling accuracy, considering the extended duration of the WRF simulation and the coastal location of Sydney [59]. The input geographical data is the moderate resolution imaging spectroradiometer (MODIS) 15s land use data. For the urban grid points in D03, we use the Sydney LCZ map generated by the LCZ Generator [60] and replace the urban areas with 11 urban categories, as shown in Figure 1 (reddish colours). The workflow about ingesting LCZ information into WRF can be found in Demuzere et al. [61]. The urban parameters used for the Sydney LCZs in WRF are obtained from Hirsch et al. [62] and Lipson et al. [63]. The details are summarised in Table 1, and all the parameters are within the ranges defined by Oke and Stewart [64].

Tables 1. Urban parameters of the LCZs in Sydney urban area in D03 for the WRF model.

| LCZ | Type | Building Height (m) Mean (std dev) | Albedo | Urban Fraction | Anthropogenic Heat (W/m$^2$) |
|---|---|---|---|---|---|
| 1 | Compact Highrise | 40.1 (18.1) | 0.107 | 0.963 | 54.0 |
| 2 | Compact Midrise | 11.6 (4.0) | 0.134 | 0.822 | 20.1 |
| 3 | Compact Lowrise | 6.2 (1.7) | 0.146 | 0.741 | 12.4 |
| 4 | Open Highrise | 21.1 (8.6) | 0.134 | 0.911 | 54.0 |
| 5 | Open Midrise | 17.5 (1.0) | 0.134 | 0.911 | 20.1 |
| 6 | Open Lowrise | 5.9 (1.7) | 0.148 | 0.673 | 10.5 |
| 7 | Lightweight Lowrise | 3.0 (1.0) | 0.160 | 0.396 | 5.4 |
| 8 | Large Lowrise | 8.8 (2.6) | 0.158 | 0.776 | 11.7 |
| 9 | Sparsely Built | 4.5 (1.5) | 0.156 | 0.269 | 3.6 |
| 10 | Heavy Industry | 9.1 (2.5) | 0.161 | 0.766 | 63.9 |
| 11 | Bare Rock or Paved | 10.0 (1.0) | 0.120 | 1.000 | 3.6 |

To assess the accuracy of the simulation results, the average hourly near-surface temperature derived from one-minute data, which is provided by the Australia Bureau of Meteorology (BOM) is employed. The statistical measures used to assess the disparity between the weather station observations and the WRF outputs include the bias, root mean square error (RMSE), and Pearson's correlation coefficient (R) [25].

The statistics from the comparison between WRF output and the observation at 8 different weather stations are summarised in Table 2. The comparison of the time series across all weather stations is provided in supplementary Figure S1. Overall, the WRF simulations are quite accurate with a bias less than 1.2°C, an R value greater than 0.8, and an RMSE around 2°C, which sufficiently characterises the dynamics of excessive urban heat in cities, as suggested by Ramamurthy et al. [65].

Table 2. Bias, RMSE and R of the simulated 2-m air temperature against in-situ observations.

| Year | Statistical Indices | Mangrove Mountain | Sydney Airport | Terrey Hills | Observatory Hill | Holsworthy Aerodrome | Sydney Harbour | Sydney Olympic Park | Horsley Park |
|---|---|---|---|---|---|---|---|---|---|
| 2017 | Bias (°C) | 0.29 | -0.26 | 0.56 | -0.69 | 0.47 | -0.89 | 0.32 | 0.22 |
|  | RMSE (°C) | 2.17 | 2.30 | 2.20 | 2.44 | 1.85 | 1.55 | 2.38 | 1.92 |
|  | R | 0.92 | 0.90 | 0.93 | 0.89 | 0.95 | 0.81 | 0.92 | 0.95 |
| 2019 | Bias (°C) | 0.67 | -0.39 | 1.06 | -0.54 | 0.54 | -0.29 | 0.56 | 0.86 |
|  | RMSE (°C) | 2.29 | 1.93 | 1.98 | 1.86 | 1.67 | 1.29 | 2.01 | 1.88 |
|  | R | 0.89 | 0.90 | 0.93 | 0.90 | 0.95 | 0.84 | 0.93 | 0.95 |
| 2020 | Bias (°C) | 0.62 | -0.11 | 1.15 | -0.34 | 0.50 | -0.63 | 0.48 | 0.38 |
|  | RMSE (°C) | 2.08 | 1.81 | 2.06 | 1.83 | 1.56 | 1.47 | 1.79 | 1.57 |
|  | R | 0.93 | 0.90 | 0.93 | 0.89 | 0.95 | 0.80 | 0.94 | 0.96 |

## 2.3 Climate Feedback-Response Analysis Method (CFRAM)

The present study aims to uncover the disparity in feedback mechanisms between non-HW and a subsequent HW of the same year and determine the primary factors. Further, it explores similarities and possible variations of the feedback mechanisms across different LCZs and over the three years containing non-HW and HW events. The definition of a HW can be found in Appendix A. CFRAM [34] assesses individual contributions of energy transport processes to the overall temperature change during HW based on energy balance over the column from surface to the atmosphere, which includes 64 atmospheric layers and a surface layer in this study. The overall workflow of using CFRAM to decompose the radiative and non-radiative processes is illustrated in Figure 2.

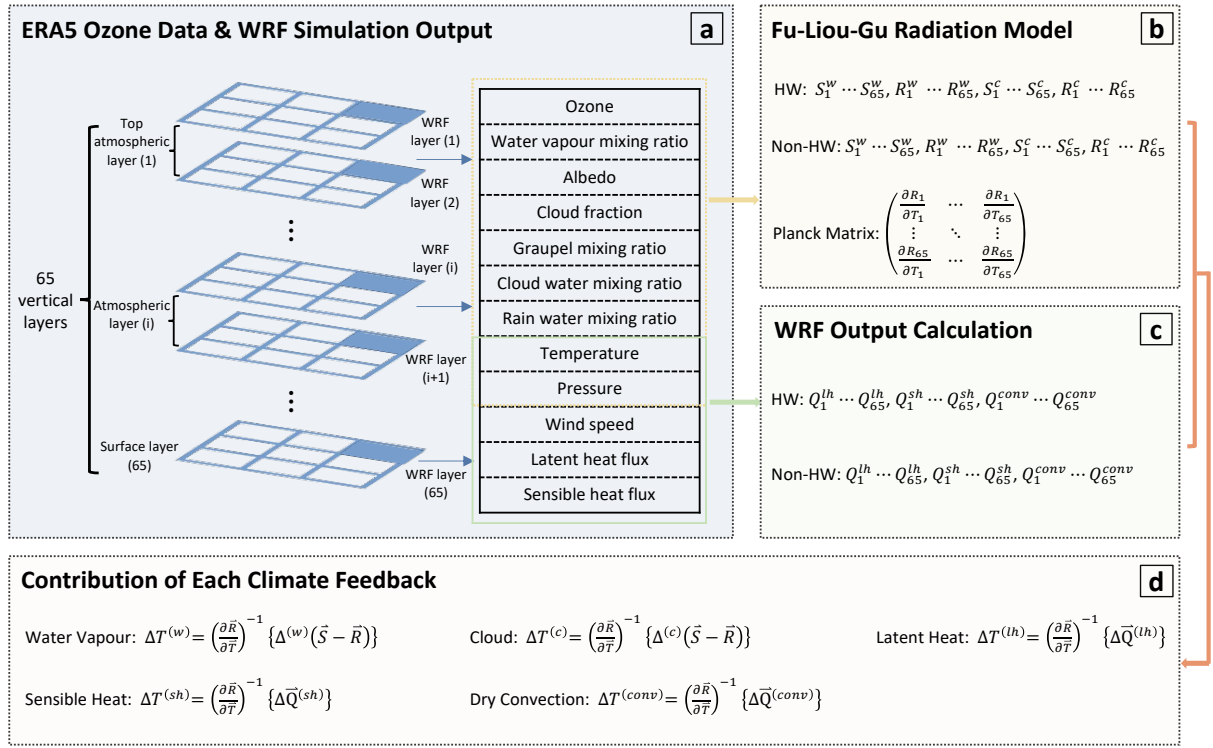

Figure 2. The overall workflow to decompose radiative and non-radiative processes using CFRAM.

The general energy balance applied in the CFRAM can be written as:

$$\Delta \frac{\partial \vec{E}}{\partial t} = \Delta \vec{S} - \Delta \vec{R} + \Delta \vec{Q}^{non-radiative} \qquad (1)$$

where $\Delta$ denotes the difference of the mean values over several days between HW and non-HW in the same year, and $\Delta \frac{\partial \vec{E}}{\partial t}$ is the difference in energy storage, which is negligible due to the interactions among various radiative feedbacks [66]. $\vec{S}$ is the convergence of (shortwave) solar radiation flux within individual layers as shown in Figure 2 (a), $\vec{R}$ is the divergence of net (longwave) infrared radiation flux within individual layers, and $\vec{Q}^{non-radiative}$ is the vertical profiles of the energy flux convergence due to non-radiative dynamic processes. $\Delta \vec{S} - \Delta \vec{R}$ can be linearised and decomposed into partial differences due to individual radiative processes as follows:

$$\Delta \vec{S} - \Delta \vec{R} \approx \Delta^{(w)}(\vec{S} - \vec{R}) + \Delta^{(c)}(\vec{S} - \vec{R}) - \left(\frac{\partial \vec{R}}{\partial \vec{T}}\right)\Delta \vec{T} \qquad (2)$$

where $w$ and $c$ represent water vapour and cloud, respectively. $\frac{\partial \vec{R}}{\partial \vec{T}}$ is the Planck feedback matrix in which the $ith$ column represents the vertical profile of changes in infrared radiative energy flux caused by a 1-K increase of the temperature in the $ith$ layer. $\Delta \vec{T}$ is the temperature difference between HW and non-HW in each layer.

$\Delta \vec{Q}^{non-radiative}$ can be separated into surface and atmospheric components including the perturbations of sensible heat flux $\Delta \vec{Q}^{(sh)}$, latent heat flux $\Delta \vec{Q}^{(lh)}$, and dry convection $\Delta \vec{Q}^{(conv)}$:

$$\Delta \vec{Q}^{non-radiative} \approx \Delta \vec{Q}^{(lh)} + \Delta \vec{Q}^{(sh)} + \Delta \vec{Q}^{(conv)} \qquad (3)$$

Substituting equations (2) and (3) into equation (1) yields:

$$\Delta \vec{T} \approx \left(\frac{\partial \vec{R}}{\partial \vec{T}}\right)^{-1} \{\Delta^{(w)}(\vec{S} - \vec{R}) + \Delta^{(c)}(\vec{S} - \vec{R}) + \Delta \vec{Q}^{(lh)} + \Delta \vec{Q}^{(sh)} + \Delta \vec{Q}^{(conv)}\} \qquad (4)$$

The Fu-Liou-Gu (FLG) radiative transfer model [67-69], which is a hybrid model combining a delta-two-four-stream approximation for infrared flux calculations [70] and a delta-four-stream approximation for solar flux calculations [71], is utilised to calculate the radiative energy differences and the Planck feedback matrix as shown in Figure 2 (b). Other parameters such as latent heat flux are obtained directly from WRF output as shown in Figure 2 (c). More details about the calculation of other non-radiative dynamic processes can be found in Appendix B.

## 3. Results and Discussions

### 3.1 Characteristics of Three Heatwave Events

Figure 3 presents the vertical variations of the potential temperature averaged over HW and non-HW periods, the vertical distribution of temperature difference between HW and non-HW periods, and temperature differences in different local climate zones in the years 2017, 2019, and 2020. The thick black lines in Figure 3 indicate the PBL heights. It is seen that the PBL height during HWs is slightly higher, with an average value of around 30 m, compared to non-HWs, due to the warmth associated with HWs, which promotes the formation of a deeper PBL. The same behaviour has also been noted in other cities, particularly during the daytime [30, 72]. Meanwhile, the deeper PBL leads to the greater effective heat capacity of the atmosphere, which in turn impacts the pattern of both surface and air temperatures. As shown in Figures 3 (a), (e) and (i), the potential temperature varies substantially during the three HW events, mostly within the PBL (below 1km), indicating that the atmosphere is more stable compared to the non-HW periods. Figures 3 (c), (g), and (k) display the vertical distributions of the temperature difference between HW and non-HW episodes. It is evident that the temperature is higher during HWs compared to non-HW periods. The maximum temperature difference may exceed 5 K, particularly during the more intense HW of 2020, when it reaches up to 9 K. The distribution of the temperature difference is complex, but a similar trend may be observed for all three HWs. The temperature difference in the atmosphere increases as one ascends from the near-surface level to around 1 km above the surface, which coincides with the normal height just above the PBL, and decreases beyond that.

To study the impact of LCZs on the temperature escalation during HWs, the vertical profiles of the mean temperature difference between HW and non-HW periods in major LCZs [60] in urban areas are plotted in Figures 3 (d), (h) and (l). An illustration of the main LCZs considered in this study is shown in Table C1 of Appendix C. LCZ is differentiated from multiple surface parameters, such as sky view factor, anthropogenic heat flux and vegetation cover. It can be seen that the variation of the temperature difference among the selected LCZs is particularly pronounced at the near-surface level since the atmospheric climate within the PBL, especially within the urban canopy layer (UCL), is strongly affected by the surface characteristics, which modify the roughness, thermal and moisture properties. At the surface, the LCZ1, LCZ2 and LCZ3, which are covered by compact highrise, compact midrise and compact lowrise, respectively, present the highest surface temperature difference during the HWs. A similar phenomenon is also noted in other cities, such as Seoul [73], where high heat stress levels are observed in LCZ1, LCZ2, and LCZ3 during the HWs. The fundamental reason for this is the significant proportion of impervious surfaces and the lower amount of vegetation in these LCZs. These factors modify the amount of incoming solar radiation, sensible heat flux and latent heat flux, resulting in relatively high surface temperature variations during HWs [74]. The high buildings also diminish convection efficiency, impacting the urban climate. Conversely, the LCZ1 and LCZ2 exhibit the least pronounced air temperature difference in the near-surface region. On the other hand, the LCZ9, characterised by sparsely built, has the most pronounced temperature difference between HW and non-HW in the near-surface region. The difference is more evident in 2020, when the HW is more intense. This is mainly because the LCZ1 and LCZ2 are predominantly situated close to the coastline, where the sea breeze modifies the high temperatures associated with the HW in these areas. Above the PBL height, the variation of the temperature difference across LCZs is negligible. These observations suggest that the land type has a substantial impact on the temperature distribution within the PBL during periods of extreme heat, particularly at the near-surface level.

Figure 4 displays the temperature variations during HWs and temperature disparities between HWs and non-HWs at the surface and 900 hPa in the years 2017, 2019 and 2020. The presence of significant temperature disparities is clearly seen in the near surface region, as shown in Figures 4 (a), (e) and (i). On the surface, distinct temperature gradients are found across the city, with the highest temperature detected in the western urban areas and considerably lower temperatures in the regions adjacent to the ocean during all three HW periods. This is because of the sea breeze, which brings cooler air to the eastern urban areas. Meanwhile, the tall buildings near the coastline reduce the surface wind speed by increasing the surface roughness [75], which in turn reduces the heat loss in the western urban areas. Similar findings are also documented in other studies [25, 62]. The temperature distribution at the 900 hPa level is homogeneous, as depicted in Figure 4 (b), (f) and (j), indicating that the temperature is evenly blended above the PBL. Heat domes are formed under the PBL during HW, and the high-pressure system creates a thermal barrier to heat transfer between the PBL and the upper atmosphere. Similar observations have been made by Huang et al. [76], Ramamurthy et al. [77] and Zhao et al. [78]. The surface temperature differential between HWs and non-HWs also exhibits a clear gradient across the city, indicating that the western urban areas experience a greater impact from HWs. Khan et al. [79] observed similar occurrences, noting that the temperature differential between inland and coastal stations increased significantly during HWs. The temperature difference also becomes homogeneous at 900 hPa, as depicted in Figure 4 (d), (h), and (l), and becomes more prominent, which means that the temperature is less affected by the surface properties.

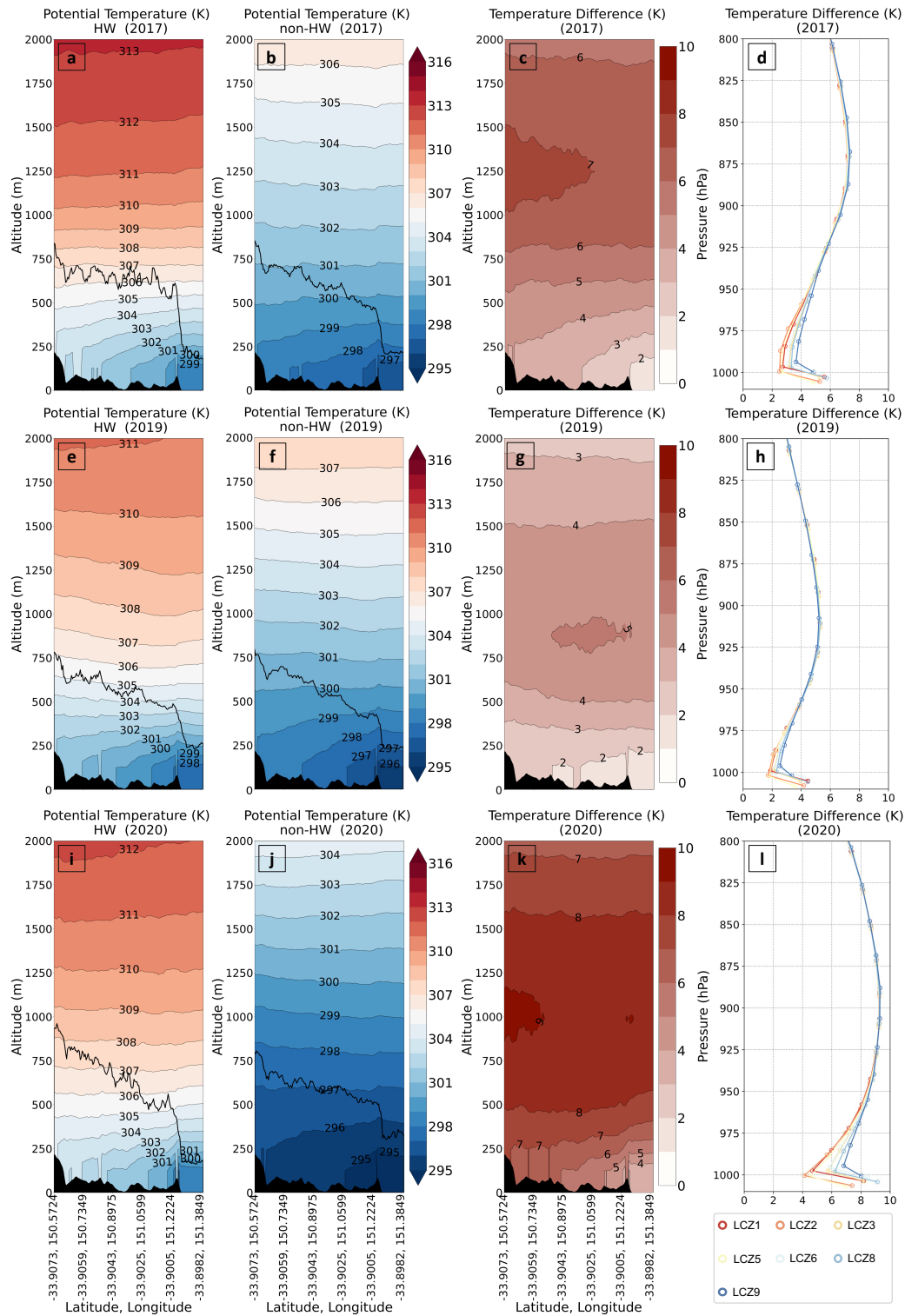

Figure 3. Vertical distribution of potential temperature during HW and non-HW, vertical distribution of temperature difference, and temperature difference in different local climate zones in (a-d) 2017, (e-h) 2019, and (i-l) 2020 (the white line refers to the PBL height).

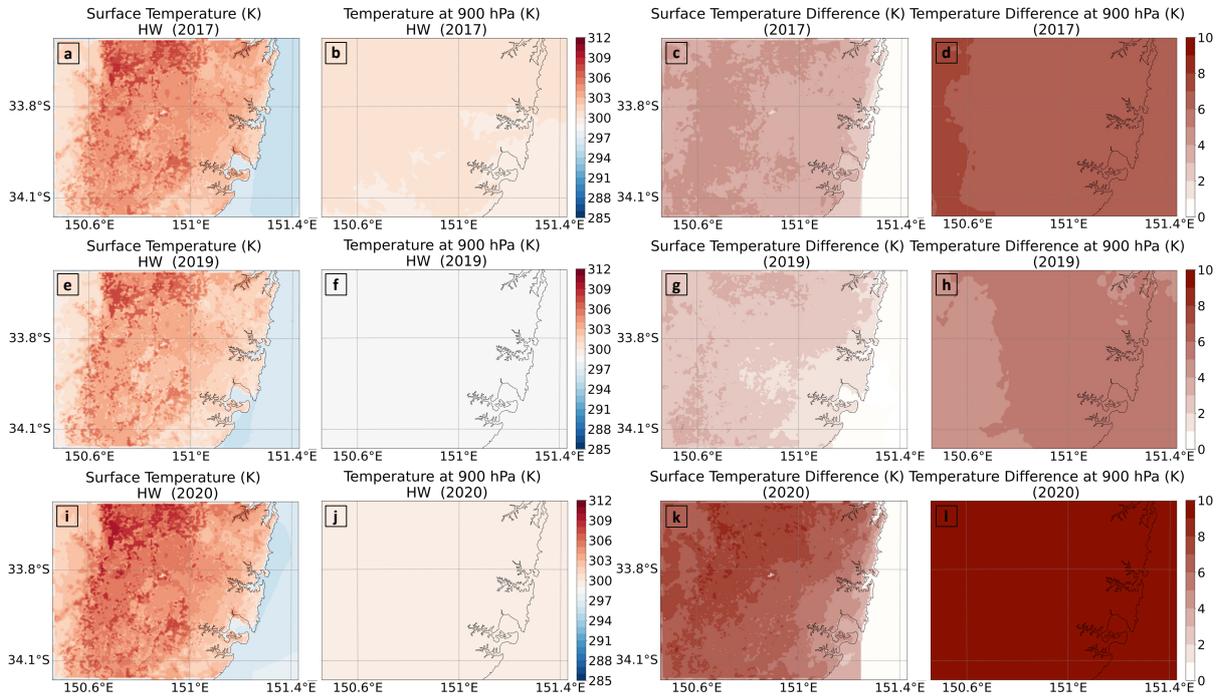

Figure 4. Temperature and temperature difference between HW and non-HW at the surface and 900 hPa in (a-d) 2017, (e-h) 2019, and (i-l) 2020.

The change in energy balance is the root cause of temperature variations between HWs and non-HWs [80, 81]. To better understand the temperature variations during HWs, we examine variation of surface energy budget between HW and non-HW periods across different LCZs in three HW events, as shown in Figure 5. During HWs, a greater amount of solar radiation reaches the surface, resulting in a noticeable rise in the surface air temperature. This effect is particularly pronounced in the year 2020, which experiences more intense and severe HWs. The mean net solar radiation difference is around 40 W/m$^2$, which is roughly twice of that in 2017 and 2019. This phenomenon is mainly due to the anomalous subsidence motions which lead to less clouds during HWs. The surface-to-atmosphere latent heat flux is also enhanced due to the increase in evapotranspiration, which is consistent with previous research of moist HWs [16]. The mean sensible and storage heat flux also increases at around 10 W/m$^2$, which alters the near-surface temperature.

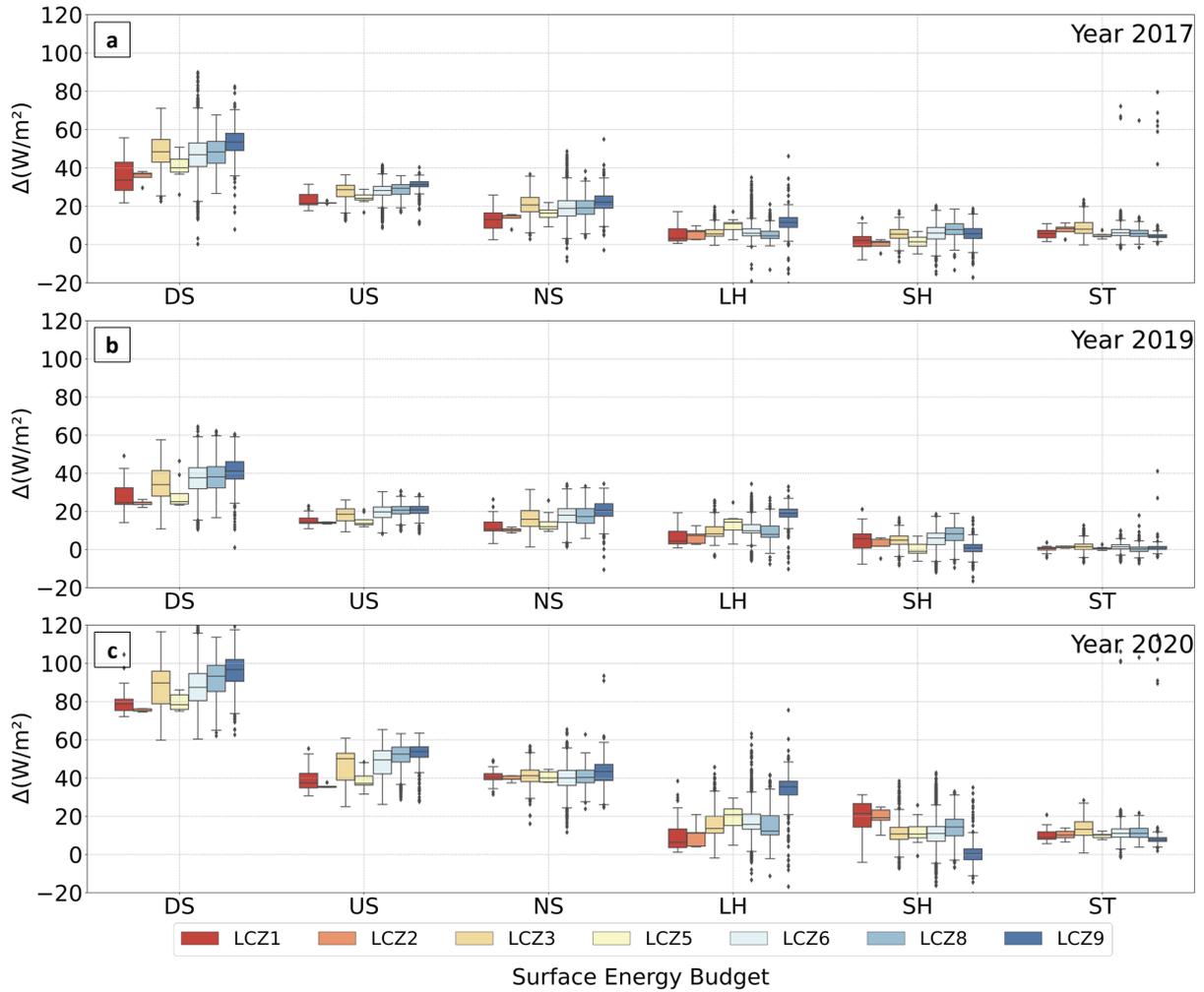

Figure 5. Surface energy budget difference between HW and non-HW periods across different LCZs in (a) 2017, (b) 2019, and (c) 2020. Anomalies of the surface energy budget include downward solar radiation (DS), upward solar radiation (US), net solar radiation (NS), latent heat flux (LH), sensible heat flux (SH), and storage heat flux (ST).

Significant disparities in incoming solar radiation are seen when comparing different LCZs. Specifically, LCZ9 receives more incoming solar radiation during HWs compared with that of non-HWs, and the difference is 1.5 times higher than those observed in LCZ1 and LCZ2. Such substantial difference also leads to a noticeable rise in latent heat flux in LCZ9 with abundant moisture. Further, the increase in latent heat flux is inversely proportional to the storage heat flux, which is less in LCZ9 compared with that in LCZ2. The differences in the surface energy budget affect the surface and near-surface temperature patterns. The overall temperature change along the PBL is further investigated in the following sections.

## 3.2 Contribution of Climate Feedback Processes in Different Local Climate Zones

Figure 6 shows the partial temperature changes generated from CFRAM for 5 individual components along the PBL. In general, the presence of water vapour gives a positive contribution to the warming of the temperature, referred to as warming effect, as shown in Figures 6 (a), (f) and (k), which is consistent with the findings in East Asia during moist HW [16]. The water vapour processes are stronger in the higher regions of the atmosphere and have a more pronounced impact on temperature anomalies occurring at the average height of PBL. In the more intense HW of 2020, the contribution of water vapour is stronger. Further, the intensity of the water vapour feedback amplifies as the temperature rises, as indicated by the Clausius-Clapeyron equation [82]. The variation of the water vapour contribution across different LCZs is very weak in 2017 and 2019, but it becomes more pronounced in 2020.

The reduction of cloud covers caused by subsidence motions associated with anomalous high-pressure systems significantly intensified the temperature anomalies during HWs. Figures 6 (b), (g) and (l) display the impact of cloud feedback for various LCZs during three HW periods. The influence of clouds on surface temperature is particularly strong since the reduction of cloud amount during HWs enhances the amount of solar radiation reaching the surface, leading to increased heat absorption. Betts et al. [83] and Wang et al. [84] also emphasized that the presence of cloud fields significantly affects both the amount of incoming shortwave and longwave radiation at the surface. These result in a monotonic rise of temperature from 14.7 to 27.4 °C and account for around 40.79% of the overall spring warming in Central Asia, respectively. The temperature gradient decreases from the surface to the near-surface area and then rises slightly. The disparity in cloud feedback across various LCZs is particularly pronounced in 2017 and 2020, when the HWs are more severe. LCZ2 and LCZ3 are less impacted by the cloud, while LCZ9 is influenced significantly by the reduction of the cloud.

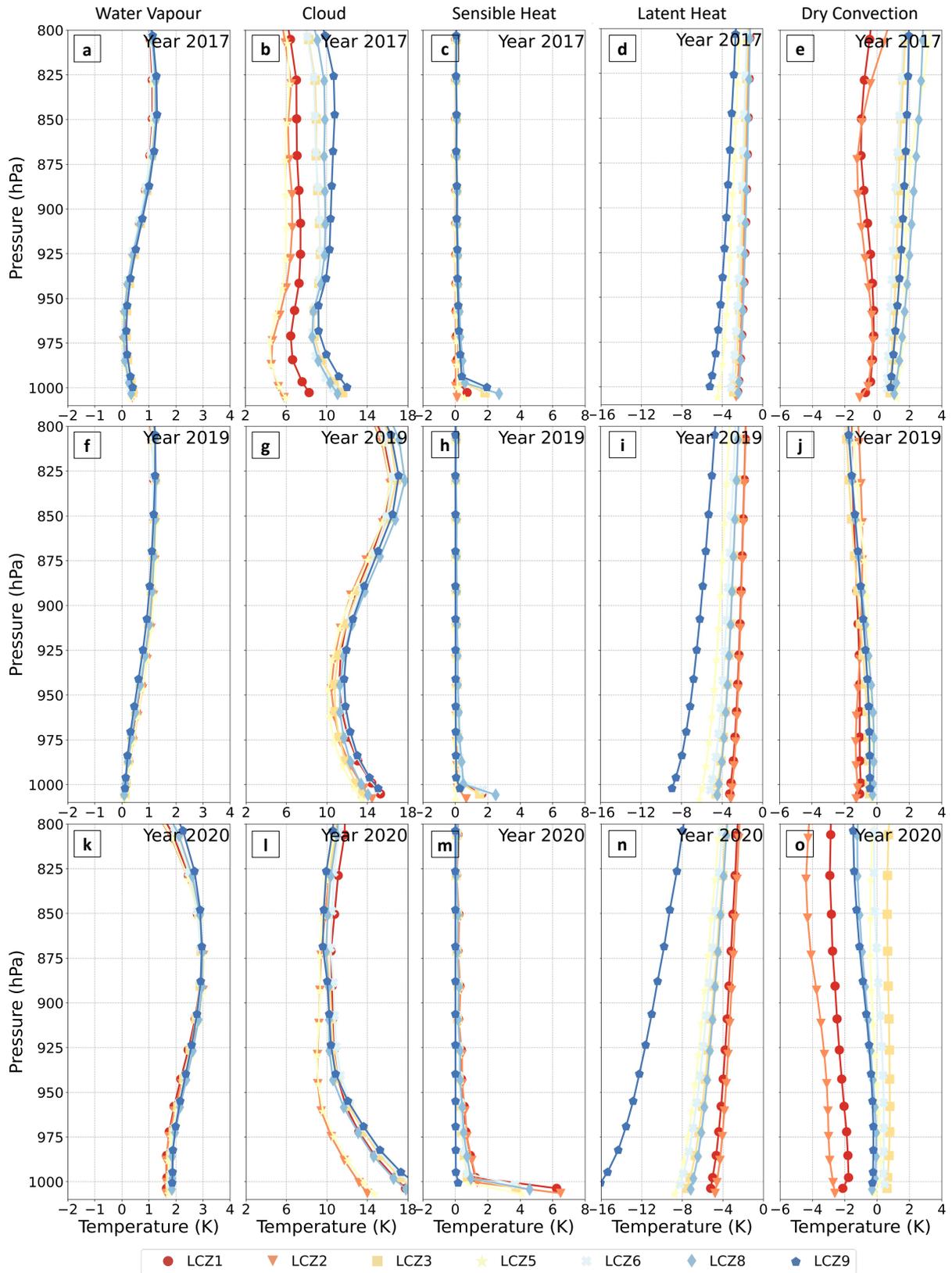

Figure 6. CFRAM-derived partial temperature anomalies along PBL due to water vapour feedback, cloud feedback, turbulent energy feedback (sensible and latent heat flux), dry

convection feedback in different LCZs in urban areas in (a - e) 2017, (f - j) 2019, and (k - o) 2020.

Turbulent energy flux comprises both sensible and latent heat fluxes (refer to Equation B5). The sensible heat flux positively contributes to the warming of the surface and near-surface temperatures by transferring heat from the surface to the atmosphere during HWs, as shown in Figures 6 (c), (h) and (m). The feedback is the strongest in 2020 as the HW is more intense. The highest temperature anomalies may reach around 7 K. Meanwhile, the contribution of the sensible heat flux to the upper atmosphere is also enhanced in 2020. For the year 2020, there is a greater increase of temperature in LCZ1 and LCZ2 compared to LCZ8 and LCZ9. However, for the year 2017, the opposite trend is observed.

In contrast to the sensible heat flux, the latent heat flux acts as a negative feedback mechanism, or cooling effect, counteracting surface and atmospheric warming by increased release of latent heat from the surface to the atmosphere, as seen in Figures 6 (d), (i) and (n). In moist HWs, the evapotranspiration of vegetation and water is enhanced due to increased incoming solar radiation and abundant soil moisture. The influence is most evident in low-density urban areas such as LCZ9. This phenomenon contrasts with the observations in dry HWs when the latent heat flux acts as positive feedback to surface warming due to deficient soil moisture [16, 24].

Figures 6 (e), (j) and (o) display the impacts of dry convection feedback in various LCZs. Typically, dry convection has a cooling effect on the overall temperature change. A greater temperature gradient, as shown in Figure 3, in the atmosphere leads to stronger vertical dry convection, which mitigates the increase of the atmospheric temperature. The phenomenon is more pronounced in high-density urban areas such as LCZ1 and LCZ2, as well as more severe HW periods. This could be attributed to the fact that most LCZ1 and LCZ2 zones are situated in close proximity to the coastline and are influenced by the coastal wind. Meanwhile, the elevated surface temperature in LCZ1 and LCZ2 results in a more pronounced buoyancy effect, hence intensifying dry convection. The disparity in dry convection contributions across various LCZs also becomes more pronounced during the 2020 HW, which is more intense.

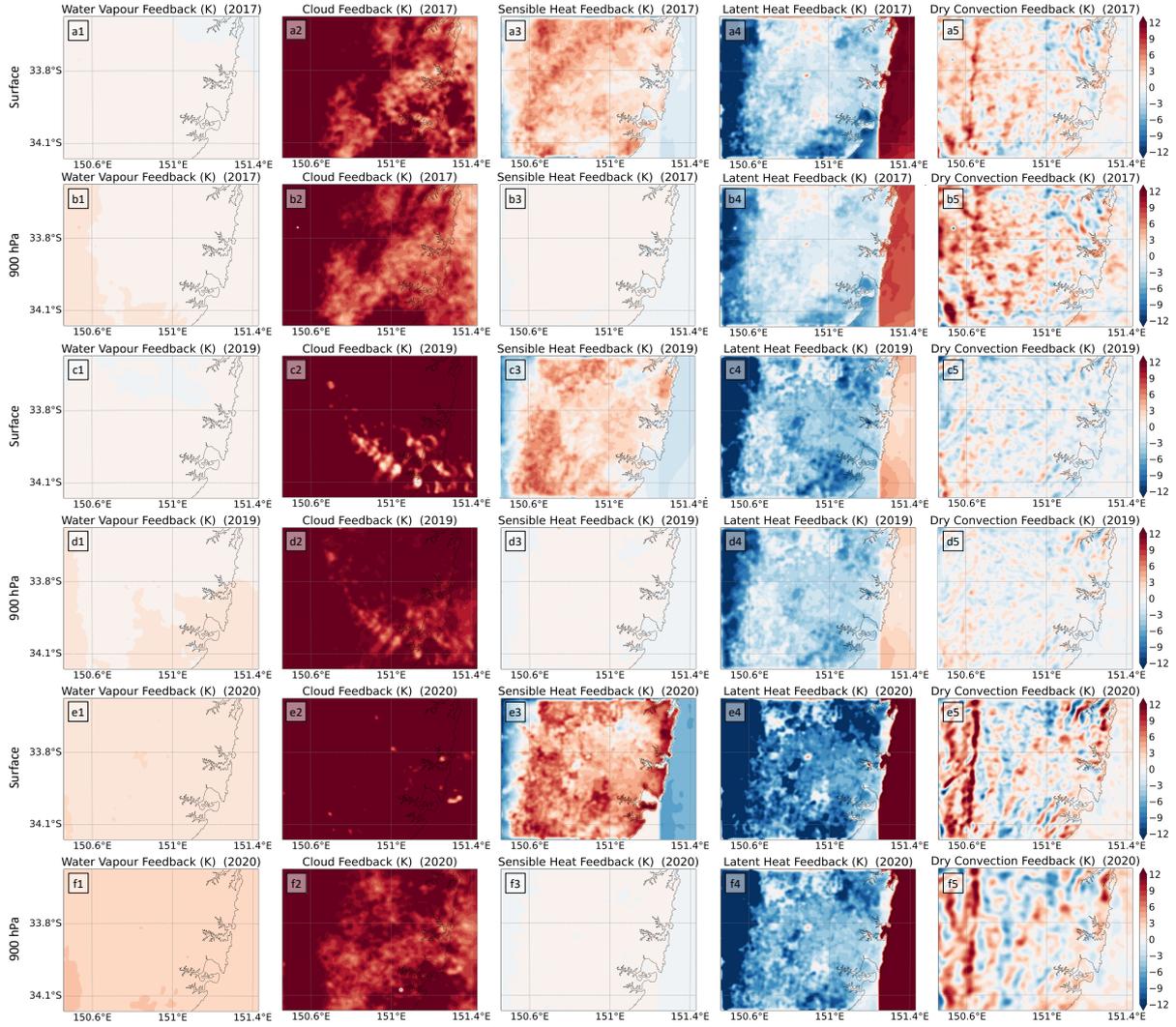

Figure 7. CFRAM-derived partial temperature anomalies due to feedbacks of water vapour, cloud, sensible heat flux, latent heat flux, and dry convection at the surface and 900 hPa in (a, b) 2017, (c, d) 2019, and (e, f) 2020.

The CFRAM-derived partial temperature anomalies due to water vapour feedback, cloud feedback, turbulent energy feedback (sensible and latent heat flux), and dry convection feedback at the surface and 900 hPa in the years of 2017, 2019 and 2020 are shown in Figure 7. The cloud feedback results in significant warming at the surface and relatively weaker warming at 900 hPa, particularly in the regions near the coast, as shown in Figures 7 (a2), (b2), (c2), (d2), (e2) and (f2). The cloud feedback near the coast region is mostly influenced by the weak shortwave cloud forcing, which is caused by a higher cloud cover in that region. This suggests that the location and distance from the coast are also significant factors affecting HW warming. The sensible and latent heat fluxes play opposite roles in temperature change in the three HWs. The sensible heat flux feedback is more pronounced at the surface, particularly in urban areas, suggesting that urbanization has a significant role in altering the sensible heat flux (Figures 7 a3, c3, and e3). The latent heat flux has a pronounced cooling impact in the rural region, and the impact is somewhat diminished at 900 hPa (Figures 7 a4, b4, c4, d4, e4, and f4). However, the increase of surface-to-atmospheric latent heat flux during three moist HWs leads to a positive water vapour feedback, which in turn adds to the warming of the surface and air (Figures 7 a1, b1, c1, d1, e1, and f1). The dry convection feedback exerts a more pronounced positive feedback effect in the western region, characterized by higher terrain compared to the eastern region and subject to hot and dry winds blowing from the west. The dry convection has a predominantly cooling impact on the coastal regions due to the coastal winds, as seen in Figures 7 (a5), (b5), (c5), (d5), (e5), and (f5).

## 3.3  Discussions

In general, the feedback of water vapour, clouds, and sensible heat flux contributes to the amplification of temperature anomalies in moist HWs. Conversely, the latent heat flux and dry convection have the opposite warming effect. In high-density urban areas, such as LCZ1 and LCZ2, sensible heat flux and cloud feedback cause significant warming anomalies during HW events. In low-density urban areas, such as LCZ8 and LCZ9, latent heat flux and clouds have more significance as opposed to high-density urban areas. In the upper atmosphere, the cloud feedback is the primary factor influencing the HW-related temperature pattern. The contributions of dry convection and latent heat flux are similar in magnitude. The presence of water vapour intensifies at higher altitudes in the atmosphere.

Our findings suggest that the LCZs have a notable influence on the near-surface temperature, particularly through modifying the sensible and latent heat fluxes. Hence, comprehending the mechanisms underlying various LCZs is crucial for assessing the efficacy of mitigation techniques for different LCZs, as not all mitigation approaches can be effectively implemented in a specific LCZ [85]. It is recommended to increase the amount of vegetation in urban areas, especially in the high-density urban districts such as LCZ1 and LCZ2 which have fewer cooling benefits from latent heat during HWs. A composite green structure consisting of shrubs, trees, and grasses, as well as the installation of rooftop and vertical greening should be considered, especially in LCZ1 and LCZ2 [86]. For low-density urban areas, such as LCZ8 and LCZ9 which receive more solar radiation during HWs, the use of reflective materials may be more beneficial in mitigating heat. Furthermore, the proximity to the coastline also has a significant impact on the temperature pattern associated with HW which influences cloud forcing.

The present work has certain limitations. CFRAM does not account for the interactive relationships among individual feedback mechanisms and temperature variations that lead to energy flux perturbations. Furthermore, this study mainly focuses on primary climate feedback and does not consider other feedback mechanisms, such as the alteration of Ozone during HWs.

## 4. Conclusions

The present study examines the characteristics of temperature and energy budget modulation between surface and atmosphere during three moist HWs in 2017, 2019, and 2020 in the Sydney area. To gain a deeper comprehension of the impact of HWs on the temperature anomalies and identify the underlying physical processes, a quantitative method called CFRAM is adopted to decompose the radiative and non-radiative feedback processes into partial temperature changes, including water vapour, sensible heat flux, latent heat flux, cloud, and dry convection.

- Our results indicate that HWs significantly impact the horizontal and vertical temperature patterns in urban areas. Heat domes are observed in the near-surface regions, and the highest temperature is detected in the western urban areas. The temperature differential between HW and non-HW may reach up to 10 K during intense HWs. Meanwhile, the potential temperature undergoes substantial fluctuations during the three HW events, with a greater temperature disparity between the surface and the atmosphere, particularly within 1 km from the surface (approximately the PBL height), suggesting a higher level of atmospheric stability.
- Regarding the influence of LCZs, the present findings indicate that the temperature anomalies among various LCZs are most noticeable at the surface and near-surface levels, especially during severe HWs when the difference of temperature anomalies among LCZs may reach up to nearly 4 K. Our findings indicate that the land cover significantly affects the temperature near the surface, and this effect is particularly noticeable during the periods of extreme HWs.
- It is also found that the reduction of cloud cover, sensible heat flux, and water vapour result in a warming effect on surface and atmosphere temperature anomalies. The reduction of cloud cover during HWs leads to increasing solar radiation, which in turn causes warming of the surface and atmospheric temperature. The phenomenon is more pronounced in remote inland areas, suggesting that the distance from the coastline also significantly influences cloud forcing. The feedback of sensible heat flux is more distinct at the surface, especially in urban regions where the contribution may reach around 6 K, indicating that urbanization plays a substantial role in modifying sensible heat flux. Conversely, latent heat flux and dry convection have contrasting influence. The latent heat flux provides the most substantial cooling to the temperature anomalies which results in a fall of surface temperature by roughly 10 K through increased release of latent heat by soil moisture to the atmosphere, and the effect is least pronounced in LCZ1 and LCZ2, which have a lower proportion of vegetation fraction.

Our findings provide a quantitative understanding of the relative contributions of radiative and non-radiative processes to temperature anomalies associated with HWs. It demonstrates that a thorough investigation, which includes numerical simulations and mathematical modelling, is necessary to gain a better understanding of the mechanisms behind HWs and the land-atmosphere interactions. In future studies, further investigation is required in diverse global regions to ascertain the primary factors that contribute to temperature anomalies in various HW categories, including moist and dry HWs. In addition, the evolution of HW on separate days prior to, during and after HW may be further studied to quantify and comprehend the fundamental mechanisms.

**Acknowledgements**

The first and last authors are grateful for the financial support of the School of Civil Engineering, the University of Sydney, and also grateful for the numerical resources provided by the National Computational Infrastructure (NCI), Australia.

**Appendix A Heatwaves in the Greater Sydney Area**

According to the BOM (Bureau of Meteorology), a HW is characterised by a duration of three or more consecutive days during which both daytime and nocturnal temperatures exhibit an exceptional increase in comparison to the prevailing long-term climatic conditions in the local area. In this study, we use the $90^{th}$ percentile criteria derived from the highest daytime temperature and minimum overnight temperature throughout the summer season (December - February) during a period of 30 years (1991 - 2020). The data is sourced from two weather stations situated in metropolitan areas, namely Sydney Observatory Hill (066062) and Sydney Airport (06037). The results are summarized in Table A1. The selected time frames for the HW events are from $9^{th}$ to $12^{th}$ February 2017, $26^{th}$ to $31^{st}$ January 2019, and $31^{st}$ January to $3^{rd}$ February 2020. The non-HW days used for comparison are 5 days before and after HW days, separately.

Table A1. Heat extremes in all weather stations.

| Date | | 2017 | | | | 2019 | | | | | | 2020 | | |
|---|---|---|---|---|---|---|---|---|---|---|---|---|---|---|
| | | 09/02 | 10/02 | 11/02 | 12/02 | 26/01 | 27/01 | 28/01 | 29/01 | 30/01 | 31/01 | 31/01 | 01/02 | 02/02 | 03/02 |
| Daytime | 066062 | √ | √ | √ | √ | √ | √ | √ | | | | √ | √ | √ | √ |
| | 066037 | √ | √ | √ | | √ | √ | | √ | √ | √ | √ | | | |
| Nighttime | 066062 | √ | | | | √ | √ | √ | √ | √ | √ | √ | √ | √ | √ |
| | 066037 | | √ | √ | √ | √ | √ | √ | √ | √ | √ | √ | √ | | |

There are various definitions of dry and moist HWs, such as defined by equivalent temperature [19, 87], relative humidity [16], and wet-bulb temperature [88]. Here, we employ the definition used by Ha et al. [16], where dry and moist HWs are those with relative humidity below 33% and above 66%, when occurring simultaneously in more than 40% of the region surface. Figure A1 shows the 2-m relative humidity against the 2-m air temperature during 3 HW periods and it is obvious that most of the grid cells have high relative humidity which is more than 33%.

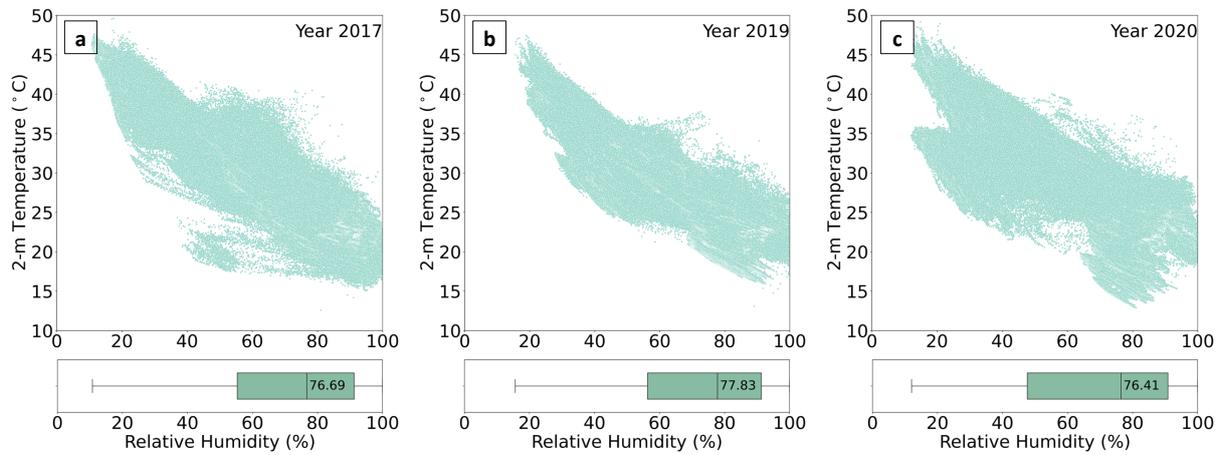

Figure A1. 2-m temperature and relative humidity during 3 HW periods.

**Appendix B Climate Feedback-Response Analysis Method (CFRAM)**

The CFRAM achieves the coupled atmosphere-surface response to individual feedback by solving the linearized infrared radiative transfer model, taking into account climate forcing and individual climate feedbacks [34]. In our study, there is a total of 65 vertical layers (as shown in Figure 2), including 64 atmospheric layers and one surface layer. The detailed calculations are shown below.

Net radiative flux of the *ith* surface/atmospheric layer:

$$(S-R)_i = \begin{cases} [(fds_i + fus_{i+1}) - (fus_i + fds_{i+1})] - [(fuir_i + fdir_{i+1}) - (fdir_i + fuir_{i+1})], & \text{atmospheric layer} \\ (fds_i - fus_i) - (fuir_i - fdir_i), & \text{surface layer} \end{cases} \quad (B1)$$

where $S_i$ is the solar radiation absorbed by the *ith* surface/atmospheric layer, $R_i$ is the infrared radiation leaving the *ith* surface/atmospheric layer, $fds_i$ is the downward solar radiation in the *ith* WRF layer, $fus_i$ is the upward solar radiation by the *ith* WRF layer, $fdir_i$ is the downward infrared radiation by the *ith* WRF layer, and $fuir_i$ is the upward infrared radiation by the *ith* WRF layer.

Planck feedback matrix:

$$\left(\frac{\partial \vec{R}}{\partial \vec{T}}\right) = \begin{pmatrix} \frac{\partial R_1}{\partial T_1} & \cdots & \frac{\partial R_1}{\partial T_{65}} \\ \vdots & \ddots & \vdots \\ \frac{\partial R_{65}}{\partial T_1} & \cdots & \frac{\partial R_{65}}{\partial T_{65}} \end{pmatrix} \quad (B2)$$

where the subscript '1' denotes the top layer of the atmosphere, '65' denotes the surface layer in our study, as shown in Figure 2 (a).

Water vapour perturbation:

$$\Delta^{(w)}(\vec{S} - \vec{R}) = \begin{pmatrix} \Delta^{(w)}(S - R)_1 \\ \vdots \\ \Delta^{(w)}(S - R)_{65} \end{pmatrix} \quad (B3)$$

Cloud perturbation:

$$\Delta^{(c)}(\vec{S} - \vec{R}) = \begin{pmatrix} \Delta^{(c)}(S - R)_1 \\ \vdots \\ \Delta^{(c)}(S - R)_{65} \end{pmatrix} \quad (B4)$$

Turbulent energy flux perturbation:

$$\Delta \vec{Q}^{turb} = \Delta \vec{Q}^{(lh)} + \Delta \vec{Q}^{(sh)} \quad (B5)$$

where $\Delta \vec{Q}^{lh}$ is the energy flux perturbation caused by the feedback from the surface latent heat flux along with the parameterized convection of the moisture that occurs simultaneously with the intended distribution of vertical heating. $\Delta \vec{Q}^{(sh)}$ is the energy flux caused by the feedback from the surface sensible heat flux.

$\Delta \vec{Q}^{(lh)}$ is given as [89]:

$$\Delta \vec{Q}^{(lh)} = \begin{pmatrix} \Delta L P_1 \\ \vdots \\ \Delta L P_{64} \\ -\Delta L E \end{pmatrix} \quad (B6)$$

where $L$ is the latent heat constant, $LE$ is surface latent heat flux, $P$ is the condensation rate at pressure level $p$. $P$ is given as:

$$\frac{dP}{dp} = \begin{cases} \frac{3E}{4\Delta p}\left[1 - \frac{(p - p_{mid})^2}{(\Delta p)^2}\right], & for\ p_{min} \leq p \leq p_{max} \\ 0, & otherwise \end{cases} \quad (B7)$$

where $E$ is the evaporation rate, $p_{min} = 292\ hPa$, $p_{max} = 831\ hPa$, $p_{mid} = (p_{min} + p_{max})/2$, and $\Delta p = (p_{max} - p_{min})/2$.

$\Delta \vec{Q}^{(sh)}$ is given as:

$$\Delta \vec{Q}^{(sh)} = \begin{pmatrix} 0 \\ \vdots \\ \Delta H \\ -\Delta H \end{pmatrix} \quad (B8)$$

where $H$ is the surface sensible heat flux.

Dry convection perturbation:

$$\Delta \vec{Q}^{(conv)} = \begin{pmatrix} (\rho C_p w \Delta T)_1 \\ \vdots \\ (\rho C_p w \Delta T)_{64} \\ (\rho C_p w \Delta T)_{65} \end{pmatrix} \quad (B9)$$

where $\rho$ is the air density, $C_p$ is the specific heat capacity, $w$ is the wind speed in the vertical direction, and $\Delta T$ is the temperature difference.

**Appendix C Local Climate Zones**

Table C1. Illustration of local climate zones discussed in this study, modified from [64] (Geographical Image Source: Google Earth).

| Built Types | Definition | Geographical Characteristics in Sydney |
|---|---|---|
| LCZ1 Compact high-rise 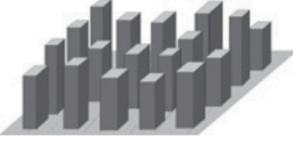 | Dense mix of tall buildings to tens of stories. Few or no trees. Land cover mostly paved. Concrete, steel, stone, and glass construction materials. | 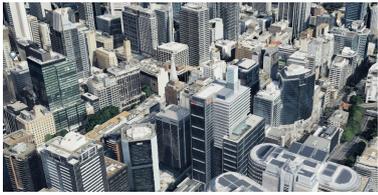 |
| LCZ2 Compact midrise 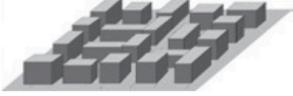 | Dense mix of midrise buildings (3-9 stories). Few or no trees. Land cover mostly paved. Stone, brick, tile, and concrete construction materials. | 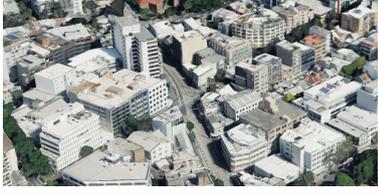 |
| LCZ3 Compact low-rise 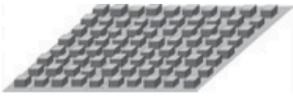 | Dense mix of low-rise buildings (1-3 stories). Few or no trees. Land cover mostly paved. Stone, brick, tile, and concrete construction materials. | 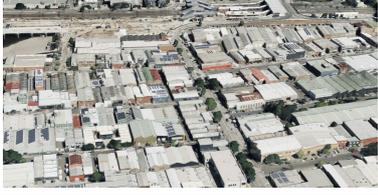 |
| LCZ5 Open midrise 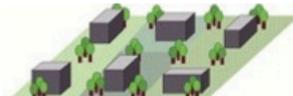 | Open arrangement of midrise buildings (3-9 stories). Abundance of pervious land cover (low plants, scattered trees). Concrete, steel, stone, and glass construction materials. | 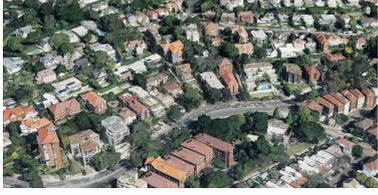 |
| LCZ6 Open low-rise | Open arrangement of low-rise buildings (1-3 stories). Abundance of pervious land cover (low plants, scattered trees). Wood, brick, stone, tile, and concrete construction materials. | |
| LCZ8 Large low-rise | Open arrangement of large low-rise buildings (1-3 stories). Few or no trees. Land cover mostly paved. Steel, concrete, metal, and stone construction materials. | |
| LCZ9 Sparsely built | Sparse arrangement of small or medium-sized buildings in a natural setting. Abundance of pervious land cover (low plants, scattered trees). | |

Supplementary Material

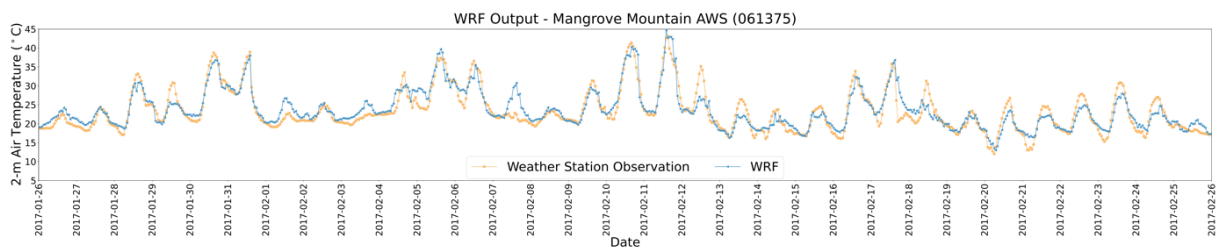
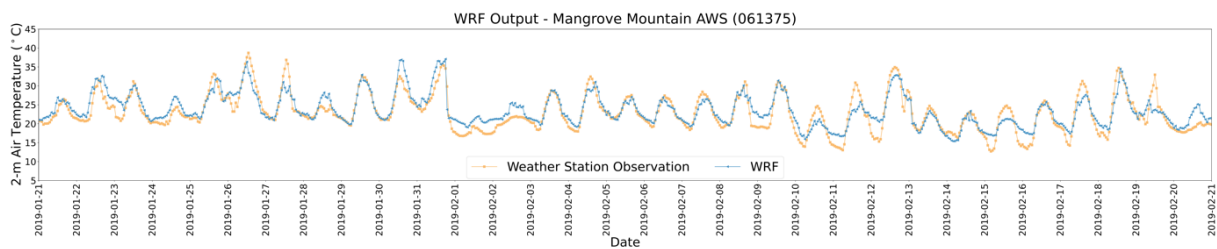
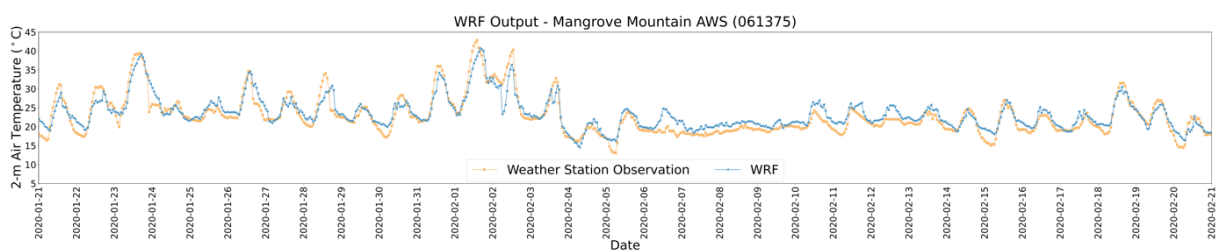
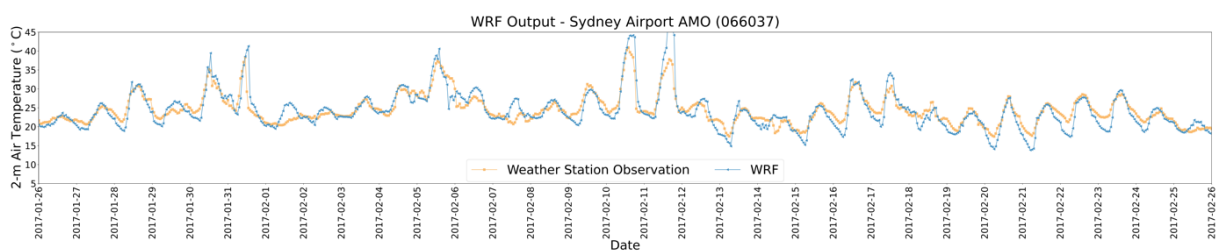
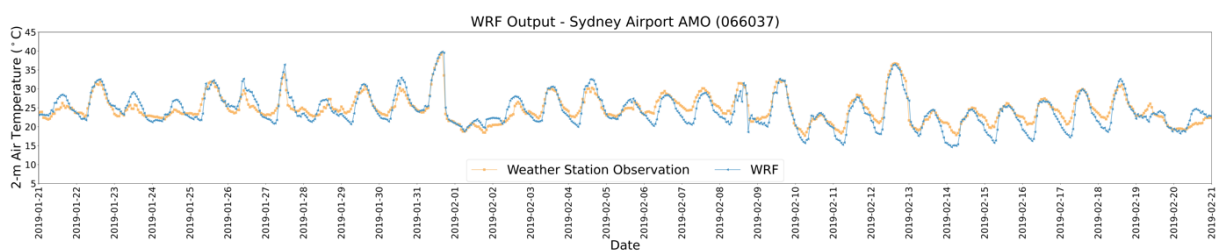
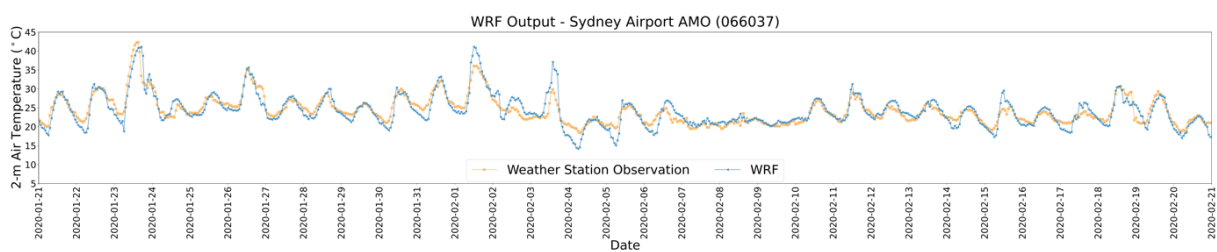
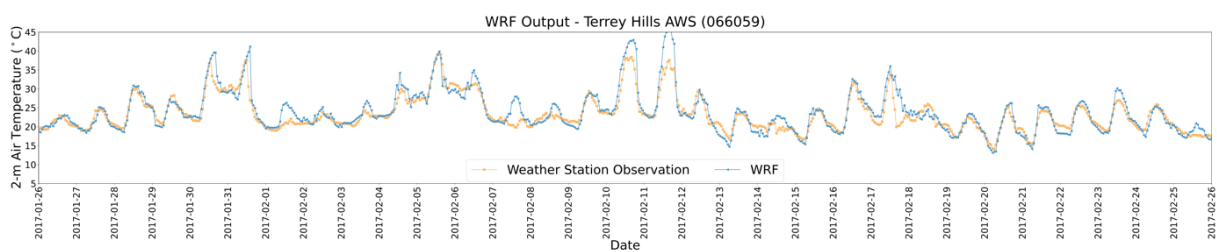

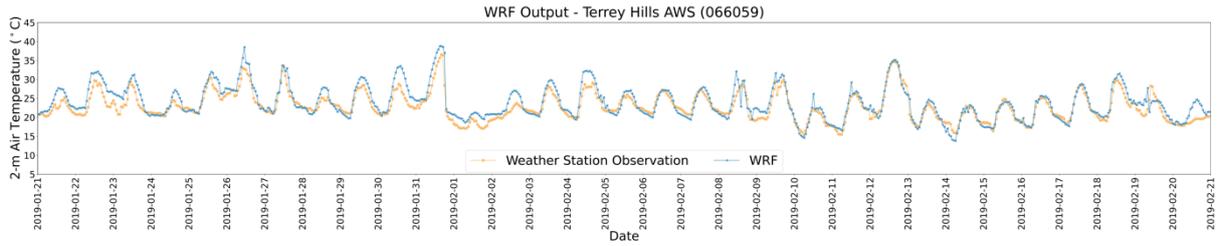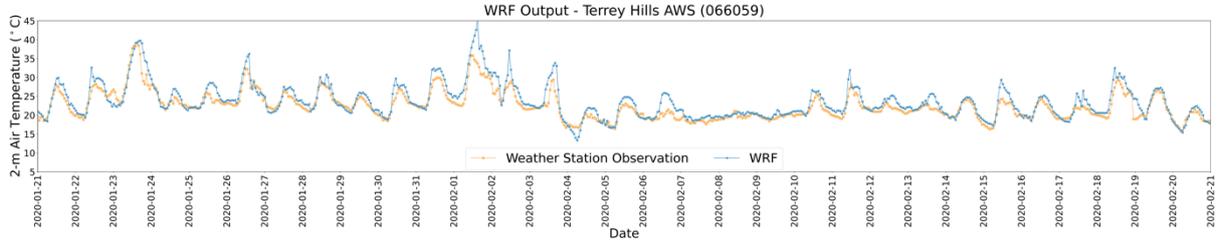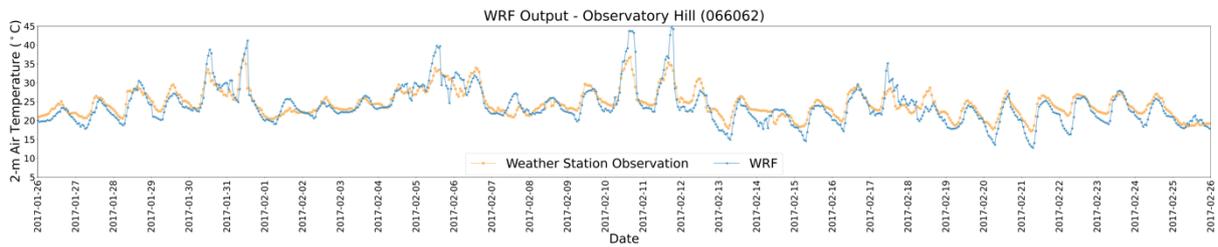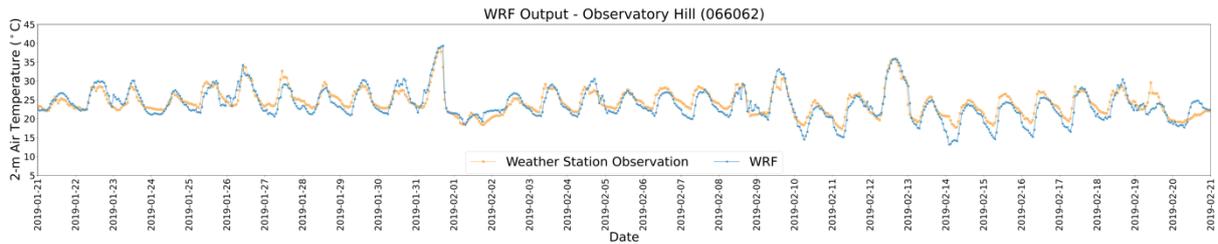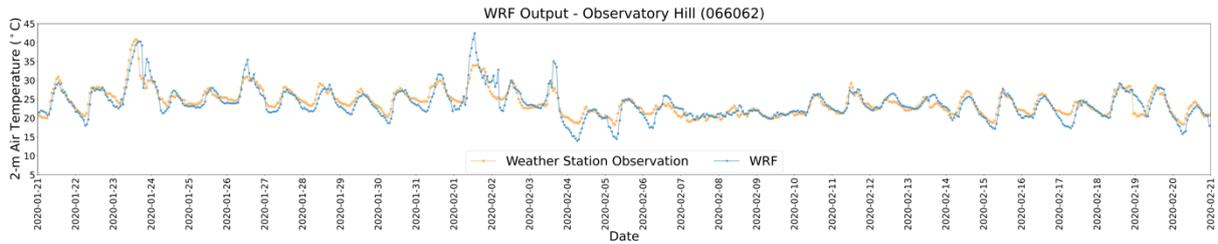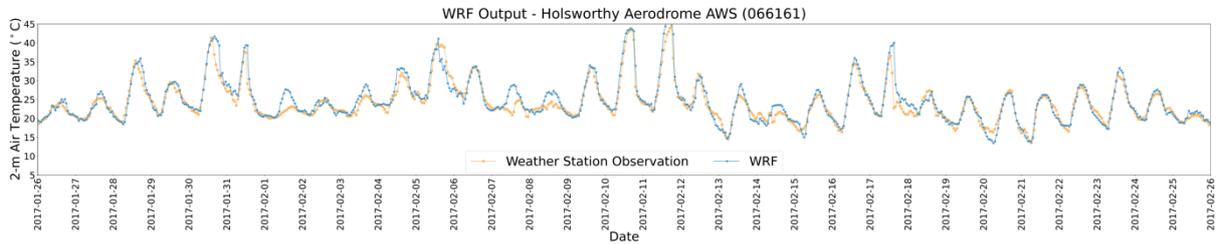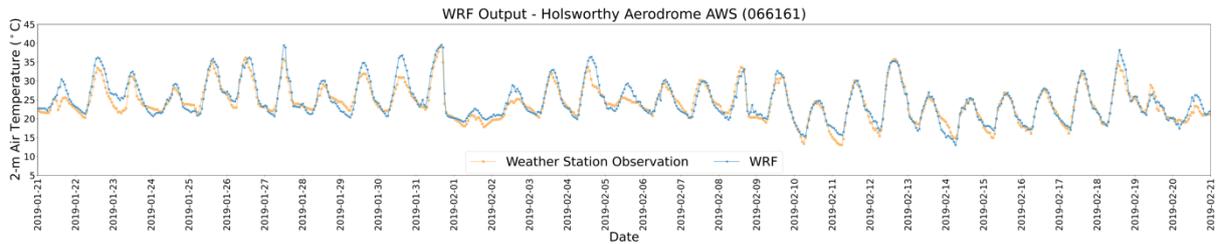

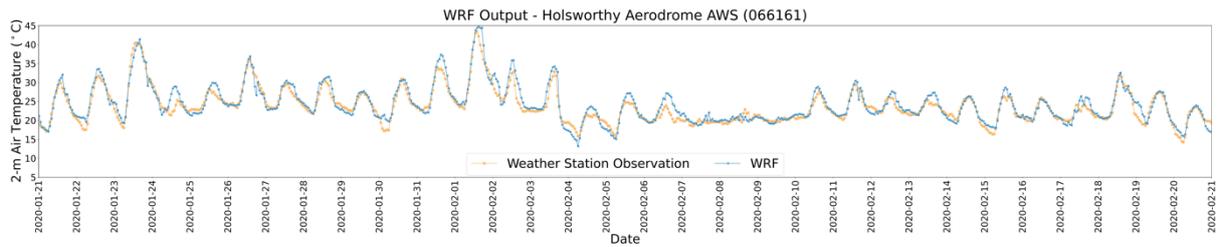
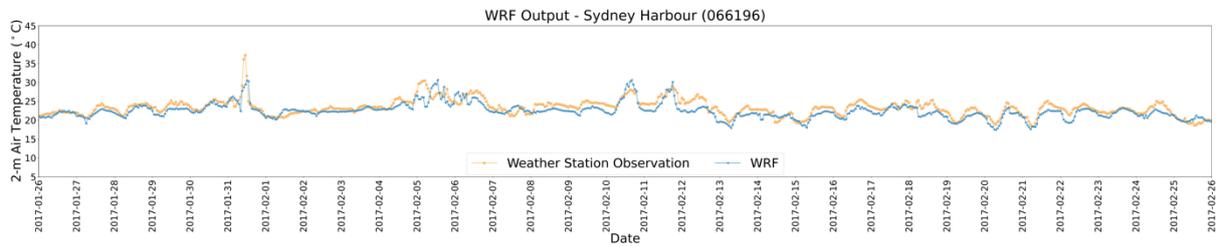
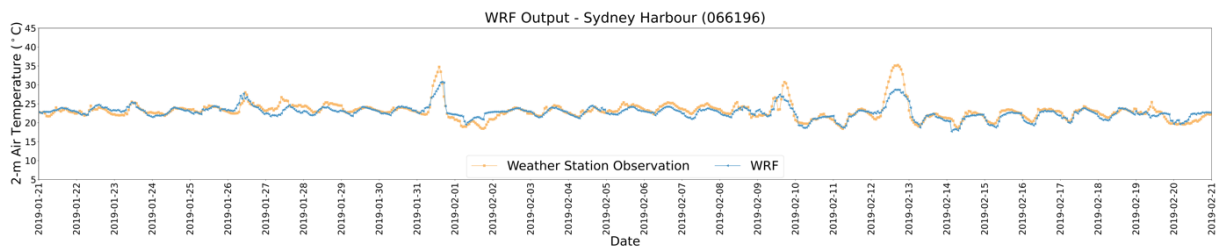
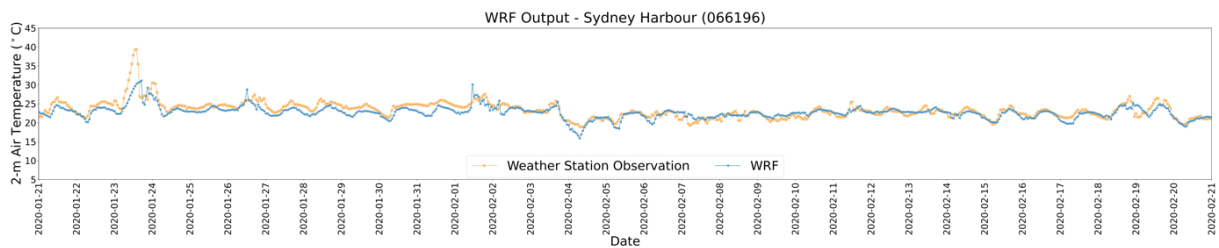
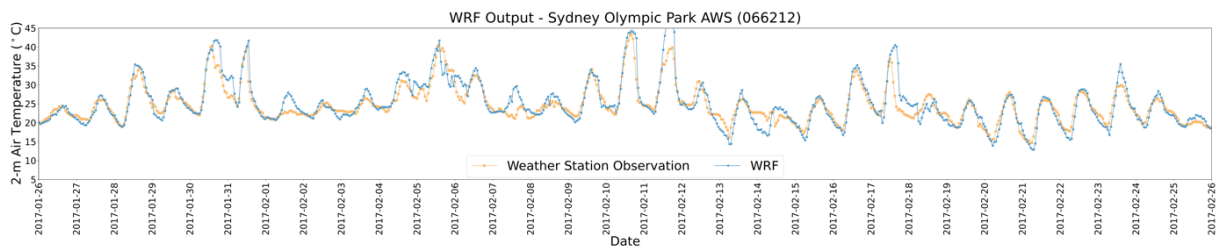
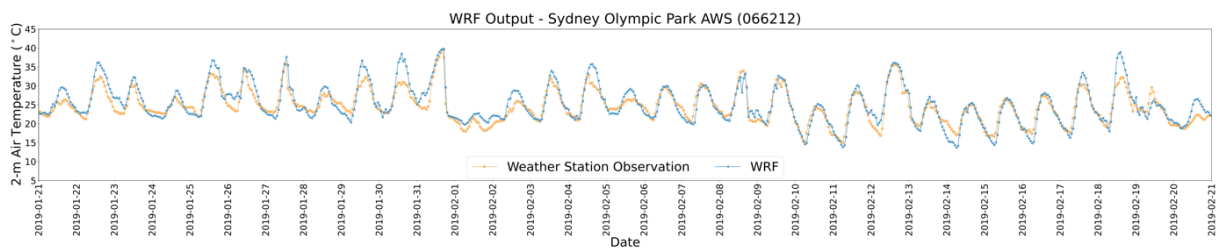
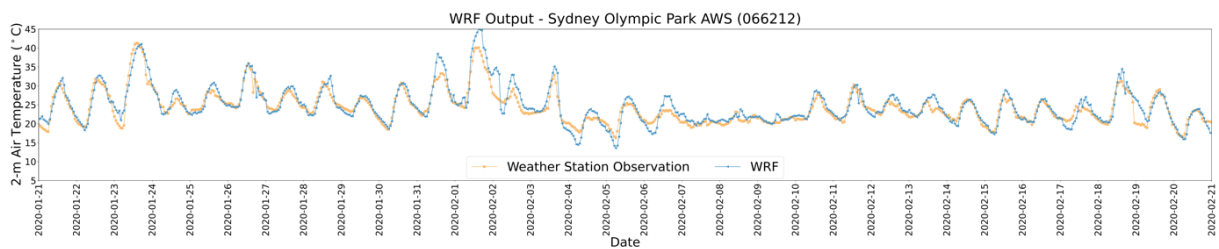

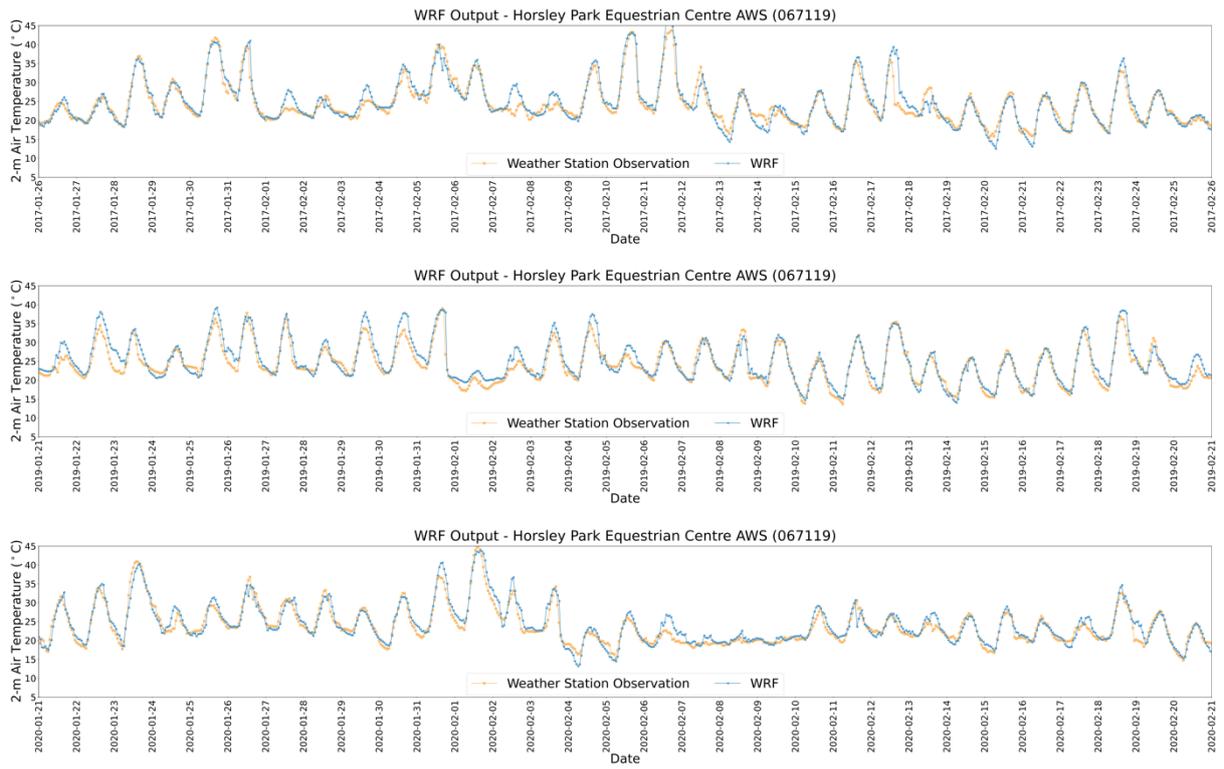

Figure S1. Comparison of results between WRF simulation and weather station observation.